\documentclass[aps,pra,groupedaddress,twocolumn,showpacs]{revtex4}
\usepackage{graphicx}
\usepackage{pstricks}

\newcommand{\ket}[1]{\ensuremath{|#1\rangle}}

\begin{document}
\title{\bf Coherent Patterning of Matter Waves with Subwavelength Localization}
\author{J. Mompart,$^{1}$ V. Ahufinger,$^{1,2} $ and G. Birkl.$^{3}$}

\begin{abstract}

We propose the Subwavelength Localization via Adiabatic Passage (SLAP) technique to coherently achieve state-selective patterning of matter waves well beyond the diffraction limit. The SLAP technique consists in coupling two partially overlapping and spatially structured laser fields to three internal levels of the matter wave yielding state-selective localization at those positions where the adiabatic passage process does not occur. We show that by means of this technique matter wave localization down to the single nanometer scale can be achieved. We analyze in detail the potential implementation of the SLAP technique for nano-lithography with an atomic beam of metastable Ne$^*$ and for coherent patterning of a two-component $^{87}$Rb Bose-Einstein condensate.

\end{abstract}

\affiliation{$^1$Departament de F\'{\i}sica, 
Universitat Aut\`{o}noma de Barcelona, 08193 Bellaterra, Spain } 
\affiliation{$^2$ICREA - Instituci\'{o} Catalana de Recerca i Estudis
Avan\c{c}ats, Barcelona, Spain } 
\affiliation{$^3$Institut f\"ur Angewandte Physik, Technische Universit\"at Darmstadt,
Schlossgartenstra\ss e 7, 64289 Darmstadt, Germany}

\date{\today }
\maketitle

\section{Introduction}

The highly controlled manipulation of atomic matter waves has proven to be an exciting field of research in recent years. Specially, research in Bose-Einstein condensation (BEC), Fermi quantum degeneracy, and quantum information processing with ultracold atoms has achieved tremendous advances \cite{Blo05}. Future progress in this field will strongly profit from optical addressability, localization, and patterning of atomic systems with a resolution not limited by the wavelength of the radiation involved. Some important examples are site-specific addressing of ultracold atoms in optical lattices \cite{addressing}, patterning of BECs \cite{bec_pat}, and atom lithography \cite{Mac07} based on light forces \cite{exp_litdip}, optical quenching \cite{exp_litmask}, or multi-photon processes \cite{teo_qualit}.

Recently, there have been several proposals for subwavelength atom localization based on the interaction of three-level atoms with light having a space-dependent amplitude distribution, mainly standing wave (SW) fields \cite{Hol96,Pas01,Sah05,Aga06,Kif08,Gor08}. In all these proposals a spatially modulated dark state is created by means of either electromagnetically induced transparency (EIT) or coherent population trapping (CPT) \cite{coh}. In fact, a proof-of-principle experiment based on the CPT technique reported intensity patterns in the transmission of a probe field presenting subwavelength spatial resolution \cite{Scu08}. Significant for the present work, the CPT technique with a SW control field produces atom localization in one of the ground states with a spatial fringe pattern ressembling that of a Fabry-Perot resonator with cavity finesse given by the ratio $\mathcal{R}$ between the control and probe field intensities \cite{Aga06}.  

\begin{figure}[tbp]
\includegraphics[width=1.0\linewidth]{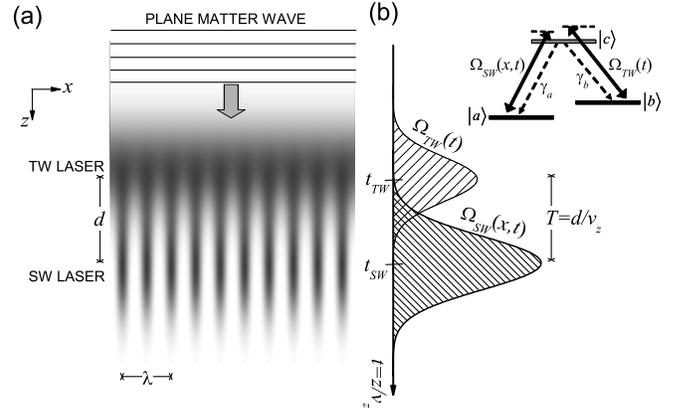}
\caption{\label{fig:scheme} (a) Schematics of the SLAP technique: A plane matter wave propagates consecutively through a TW and a partially overlapping SW laser field either in space (as shown here) or in time. (b) Three-level atomic system and Gaussian temporal profiles of the Rabi frequencies $\Omega_{TW} (t)$ and $\Omega_{SW} (x,t)$. $\gamma_a$ and $\gamma_b$ account for the spontaneous emission decay rates of the corresponding transition.}
\end{figure}
In this paper, we propose a state-selective atom localization and patterning scheme based on Stimulated Raman Adiabatic Passage (STIRAP) \cite{Ber98,Kuk89} that, compared to the CPT based techniques, presents several important advantages: (i) it produces 'super-localization', i.e.,  narrower localization than that expected from the CPT-finesse parameter $\mathcal{R}$; (ii) it is a fully coherent process that does not rely on spontaneous emission to the dark state and, therefore, it can be applied to open three-level systems and to systems where coherence has to be preserved such as BECs; (iii) the localized state does not suffer from recoil induced broadening and, therefore, the Raman-Nath approximation holds \cite{RNM01}, and, finally, (iv) it is robust under uncontrolled variations of the system parameters, e.g., intensity fluctuations of the laser fields. We describe here the main features of this Subwavelength Localization via Adiabatic Passage (SLAP) technique, as well as its potential implementation for matter wave lithography down to the single nanometer scale and for coherent patterning of a BEC at the Heisenberg limit. 
Note that STIRAP without the spatial localization feature introduced here has been proposed \cite{molecularBEC_th} and recently experimentally demonstrated \cite{molecularBEC_exp} for the transition from an atomic to a molecular BEC and for the optical control of the internal and external angular momenta of an extended BEC \cite{vortexBEC}. 

The paper is organized as follows. In Section II we describe the basics of the SLAP technique and derive semi-analytical conditions for achieving the 'super-localization' regime. In Section III and IV we discuss the application of the SLAP technique for nano-lithography with a Ne* atomic beam and for coherent patterning of a two-component $^{87}$Rb BEC, respectively. In section V we further comment on other possible applications of the SLAP technique and present a short conclusion.    

\section{SLAP technique}

The schematics of the SLAP technique are shown in Fig.~1. A plane matter wave formed by three-level atoms in a $\Lambda$-type configuration propagates at a velocity $v_z$ through two partially overlapping laser fields: the traveling wave (TW) couples the $\ket{c} \leftrightarrow \ket{b}$ transition with a Rabi frequency $\Omega_{TW}(t) = \Omega_{TW0} \exp(-(t-t_{TW})^2/\sigma_{TW}^2)$ and the SW couples the $\ket{c} \leftrightarrow \ket{a}$ transition with a Rabi frequency $\Omega_{SW} (x,t) = \Omega_{SW0} \sin{kx} \exp(-(t-t_{SW})^2/\sigma_{SW}^2)$. $k=2\pi / \lambda$ is the SW field wave number and $T= t_{SW}-t_{TW}= d/{v_z}$ the characteristic STIRAP time with $d$ the spatial separation between the centers of the two laser beams. $\Delta_{TW}=\omega_{TW}-\omega_{cb}$ ($\Delta_{SW}=\omega_{SW}-\omega_{ca}$) is the single-photon detuning between the TW (SW) field and the corresponding transition. $\gamma_a $ ($\gamma_b$) is the spontaneous emission decay rate from $\ket{c}$ to $\ket{a}$ (from $\ket{c}$ to $\ket{b}$). The spatial and the temporal variants of the SLAP technique are connected by the simple transformation $t=z/v_z$.

Under the two-photon resonance condition $\Delta_{TW}=$ $\Delta_{SW}$, one of the position-dependent energy eigenstates of the $\Lambda$-type three-level system is the so-called dark state $\ket{D(x,t)}=\cos\theta (x,t) \ket{a} - \sin \theta (x,t) \ket{b}$ where $\tan \theta (x,t) = 
\Omega_{SW}(x,t) / \Omega_{TW} (t)$. STIRAP \cite{Ber98} consists in following this energy eigenstate from $\ket{\psi_{in}} = \ket{a}$ to $\ket{\psi_{out}} =\ket{b}$ smoothly changing $\theta$ from $0^0$ to $90^0$ by means of two partially overlapping laser fields as in the counterintuitive sequence of Fig.~1. To keep the system in the energy eigenstate, this process must be performed fulfilling the 'global' adiabaticity condition \cite{Ber98}:
\begin{equation}
\Omega_{SW0}^2\sin^2{kx}+\Omega_{TW0}^2 > \left( {A \over T } \right)^2 \\
\end{equation}
where $A$ is a dimensionless constant that for optimal Gaussian profiles and overlapping times takes values around 10 \cite{Kuk89}.

In the SLAP technique, we assume that the matter wave has been initially prepared, by means of e.g., optical pumping, into the internal state $\ket{a}$. Then, those atoms crossing the nodes of the SW remain in state $\ket{a}$ while those interacting with the TW and the SW fields through the STIRAP process are transferred to state $\ket{b}$. Therefore, an intense SW field should produce sharp peaks on the spatial population distribution of state $\ket{a}$ at its nodes. From Eq.~(1) and assuming $A> T \Omega_{TW0}$, the FWHM of these peaks is given by:
\begin{equation}
\left( \Delta x \right)_{\rm SLAP} =  \left( \Delta x \right)_{\rm CPT} {1 \over 2} \sqrt{\left(A \over {T \Omega_{TW0}}\right)^2-1} \\
\end{equation}
where $\left( \Delta x \right)_{\rm CPT} = 2 / k \sqrt{\mathcal{R}}$ with $\mathcal{R} \equiv \Omega_{SW0}^2 / \Omega_{TW0}^2 $ is the FWHM of the peaks in the Fabry-Perot type localization that would be attained by means of the CPT technique \cite{Aga06}. Therefore, for
\begin{equation}
T \Omega_{TW0} = {d \over v_z} \Omega_{TW0} > {A \over \sqrt{5}}\\
\end{equation}
the 'super-localization' regime which we define by $\left( \Delta x \right)_{\rm SLAP} < \left( \Delta x \right)_{\rm CPT}$ is reached. Note that for $A=10$ corresponding to optimal parameter values \cite{Kuk89}, condition (3) reads $T \Omega_{TW0} > 4.5 $. 

\begin{figure}[tbp]
\includegraphics[width=1.0\linewidth]{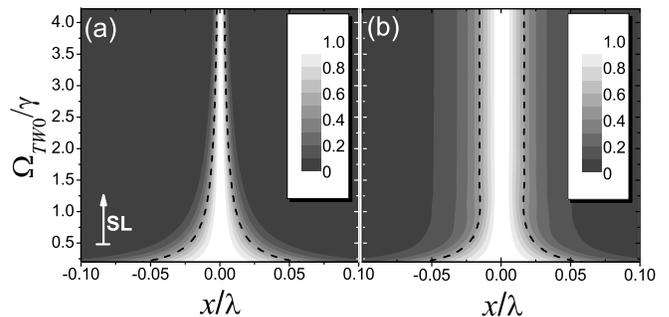}
\caption{\label{fig:cptstirap} Population distribution of state $\ket{a}$ after applying (a) the SLAP technique and (b) the CPT technique as a function of the amplitude of the TW Rabi frequency $\Omega_{TW0}$ for $\mathcal{R}=100$, $\gamma \sigma_{TW} = \gamma \sigma_{SW} = 5 $, $\Delta_{TW}=\Delta_{SW}=0$, $\gamma T=10 $ for the SLAP case, and $\gamma T=0$ for the CPT case. The horizontal separation between the dashed curves gives the FWHM of the corresponding localized structure. SL in (a) indicates the regime of 'super-localization' with $\left( \Delta x \right)_{\rm SLAP} < $ $\left( \Delta x \right)_{\rm CPT}$.}
\end{figure}

Fig. 2 shows numerical simulations of the state selective localization process by integrating the corresponding density matrix equations for both the SLAP and the CPT techniques. In the CPT technique \cite{Aga06}, subwavelength state-selective localization is obtained by reaching the steady-state through an optical-pumping process to the dark-state involving several cycles of laser excitation and spontaneous emission. In the setup of Fig.~1, the CPT process corresponds to $T=0$, $\sigma_{TW}=\sigma_{SW}$ and $\gamma \sigma_{TW} \gg 1$, where we have assumed, for simplicity, $\gamma_a = \gamma_b (\equiv \gamma)$. For $\mathcal{R}=100$ and the rest of parameters given in the figure caption, Fig.~2(a) shows that for $\Omega_{TW0} > 0.45\gamma $ the super-localization condition (3) is fulfilled and the SLAP technique yields better localization than the CPT technique, i.e., $\left( \Delta x \right)_{\rm SLAP} < $ $\left( \Delta x \right)_{\rm CPT} \sim 0.032 \lambda$. 

Note that we have considered here, for simplicity, a 1D SW field in the $\ket{a}$-$\ket{b}$ transition, although the SLAP techniques applies also to higher dimensions and to any arbitrarely spatially structured field presenting intensity nodes.

\section{SLAP based nano-lithography}

As a first implementation, we consider atom lithography based on substrates sensitive to the internal energy of metastable atoms \cite{ESL99}. For this purpose, we take a plane matter wave of metastable Ne* whose initial internal level $=2p^53s(^3P_0)$ has an energy of 16.6~eV and thus high potential for surface damage. In fact, Ne* is a prime candidate for coherent manipulation, since the STIRAP technique has been successfully reported with Ne* \cite{MSB96} using the $\Lambda$ scheme $2p^53s(^3P_0) \leftrightarrow 2p^53p(^3P_1) \leftrightarrow 2p^53s(^3P_2)$ where the first and the last are long lived states (see Fig.~3(a)). However, here we are interested in applying the SLAP technique such that, away from  the nodes of the SW, the initial state is adiabatically transferred to a fast decaying state in order to remove the corresponding high internal energy. Thus we consider the open three-level $\Lambda$ scheme $2p^53s(^3P_0) \leftrightarrow 2p^53p(^3P_1) \leftrightarrow 2p^53s(^3P_1)$ (depicted in Fig.~3(a)) with state $2p^53s(^3P_1)$ decaying to the ground state at a rate of $2 \pi \times 7.58 \cdot 10^6 {\, \rm s}^{-1}$. Fig.~3(b) shows subwavelength atom localization in state $2p^53s(^3P_0)$ (solid curve) around a node of the SW (period of 308.2~nm) after the application of the SLAP technique for $\mathcal{R}=400$ and the other parameters given in the figure caption. Note that although part of the population (dashed curve) is diabatically transferred to the high energy state $2p^53s(^3P_2)$  with lifetime $\tau =14.73 {\, \rm s}$ \cite{NeGerhard03}, this population could be efficiently pumped to the ground state via $2p^53p(^3D_2)$ with an extra laser field. Thus, for realistic parameter values one would expect state selective localization with a FWHM of only a few ${\rm nm}$ yielding high contrast peak energies of 80\% (solid curve in Fig.~3(c)) in the absence of depumping of the $2p^53s(^3P_2)$ state, and of nearly 100\% (dashed curve in Fig.~3(c)) in the presence of the depumping.

As an important feature of the SLAP technique, the localized state $\ket{a} $ does not interact with the light fields at any time and therefore does not suffer from recoil induced broadening, which implies that the Raman-Nath approximation perfectly applies \cite{RNM01}.
In this situation, the transversal velocity spread of the initial matter wave determines the  
ultimate resolution limit of the SLAP technique. Taking $\overline{\Delta v}_x$ as the rms transversal velocity spread, the limit $T \overline{\Delta v}_x \ll {\left( \Delta x \right)_{\rm SLAP}}$ corresponds to ${\overline{\Delta v}_x / v_z } \ll {\left( \Delta x \right)_{\rm SLAP} / d}$. 
Thus, for typical parameters, $v_z = 500 {\,\rm m/s}$, $\overline{\Delta v}_x = 5 {\,\rm cm/s}$, and $d=2{\,\rm \mu m}$, localization down to single nm can be achieved. As given by the Heisenberg uncertainty principle, strong atom localization should also result in the appearance of high momentum components \cite{Aga06}. For the results shown in Fig.~3(b), we have verified that the highest momentum components do not have time enough to smear out the localized structure until the end of the SW where the substrate is placed.

\begin{figure}[tbp]
\includegraphics[width=0.9\linewidth]{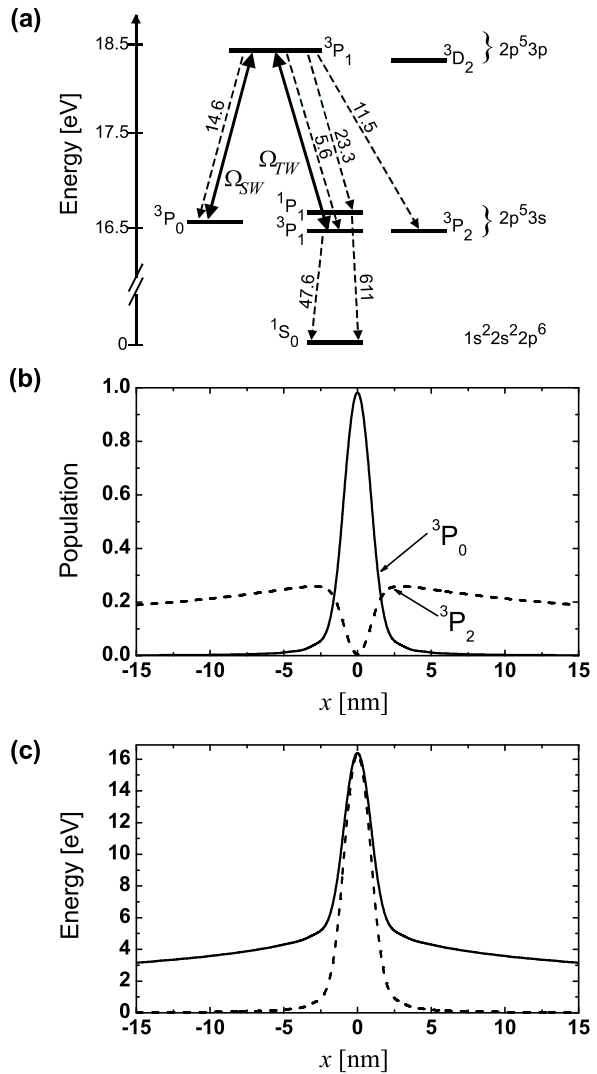}
\caption{\label{fig:cptstirap} SLAP technique for a Ne* matter wave: (a) Relevant energy levels and Einstein A coefficients (in units of $10^6 {\, \rm s}^{-1}$) for Ne*. The TW field at $\lambda_{TW}=603.0{\, \rm nm}$ couples transition 
$2p^53s(^3P_1) \leftrightarrow 2p^53p(^3P_1)$ while the SW at $\lambda_{SW}=616.4 {\, \rm nm}$  couples 
$2p^53s(^3P_0) \leftrightarrow 2p^53p(^3P_1)$. 
(b) Final spatial population distribution around a node of the SW for state $2p^53s(^3P_0)$ (solid curve) and $2p^53s(^3P_2)$ (dashed curve). (c) Spatial distribution of the Ne matter wave internal energy after the SLAP technique around a node of the SW (solid curve) and after the depumping process of state $2p^53s(^3P_2)$ (dashed curve).
Parameters: $v_z = 500 {\, \rm m/s}$, $d=100 {\, \rm \mu m}$, $\sigma_{TW}=\sigma_{SW}=100{\, \rm ns}$, $\mathcal{R}=400$, $\Omega_{TW0} = 2 \pi \times 1.6\cdot 10^7 {\, \rm s}^{-1}$, and $\Delta_{TW}=\Delta_{SW}=0$.}
\end{figure}

\section{Coherent patterning of a BEC based on SLAP}

\begin{figure}[tbp]
\includegraphics[width=1.0\linewidth]{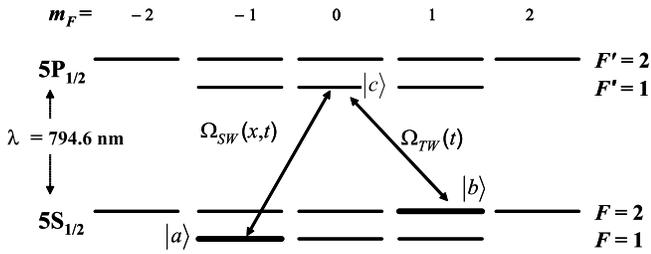}
\caption{Hyperfine structure of the D$1$ transition line of $^{87}$Rb with the couplings 
$\Omega_{SW} (x,t)$ and $\Omega_{TW} (t)$ defining the $\Lambda$-scheme. The broader lines correspond to the two condensed trapped states $\ket{a}=\ket{F=1,m_F=-1}$ and $\ket{b}=\ket{F=2,m_F=1}$.}\label{rb87}
\end{figure}

As a second implementation, we now focus on a trapped BEC of $^{87}$Rb to show the feasibility to generate narrow structures in the condensate by means of the SLAP technique. The $\Lambda $-type three level configuration under study is depicted in Fig.~\ref{rb87}. We consider a zero temperature two-species $^{87}$Rb BEC, $\ket{a}=\ket{F=1,m_F=-1}$ and $\ket{b}=\ket{F=2,m_F=1}$, confined in a one dimensional geometry. The description of the system is performed within the 1D coupled Gross-Pitaevskii equations:
\begin{eqnarray}
i\hbar \frac{d\psi_a}{dt}&=&\left[ -\frac{\hbar^{2}}{2m}\triangle +V_a(x)+g_{aa}|\psi_a  |^{2}+g_{ab}|\psi_b|^{2}\right] \psi_a \nonumber\\&+&\frac{1}{2}\hbar\Omega_{SW}(x,t) \psi_c\\
i\hbar \frac{d\psi_b}{dt}&=&\left[ -\frac{\hbar^{2}}{2m}\triangle +V_b(x)+g_{bb}|\psi_b |^{2}+g_{ab}|\psi_a|^{2}\right] \psi_b \nonumber\\&+&\frac{1}{2}\hbar\Omega_{TW}(t) \psi_c+\hbar(\Delta_{SW}-\Delta_{TW})\psi_b\\
i\hbar \frac{d\psi_c}{dt}&=&\frac{1}{2}\hbar\Omega_{SW}(x,t) \psi_a+\frac{1}{2}\hbar\Omega_{TW}(t)\psi_b \nonumber\\&-& i\frac{\Gamma}{2}\psi_c+\hbar\Delta_{SW}\psi_c 
\label{GPE} 
\end{eqnarray}
where the effective 1D nonlinearity is given by $g_{ij}= 2\hbar a_{ij}\omega _{t}$, $i,j=a,b$ with 
$a_{ij}$ the interspecies ($i\not=j$) and intraspecies ($i=j$) $s$-wave scattering lengths, and $\omega_t$ the transverse trapping frequency. In $^{87}$Rb the scattering lengths are known to be in the proportion $a_{aa}:a_{ab}:a_{bb}=1.03:1:0.97$ with the average of the three being $55(3){\rm \AA}$ \cite{scattering}. Since the magnetic moments of the two trapped components are the same to first order, magnetic trapping as well as optical trapping is possible with equal potentials for both components. The axial trapping potential reads $V_a(x)=V_b(x)= m\omega^2_x x^2/2$, with $m$ the $^{87}$Rb mass and $\omega_x$ the axial trapping frequency. State $\ket{c}=\ket{F'=1,m_F=0}$ is not trapped and the excited atoms are assumed to escape from the BEC at a rate $\Gamma = 2\pi \times 5.41\cdot 10^6{\rm s}^{-1}$.

\begin{figure}[tbp]
\includegraphics[width=0.9\linewidth]{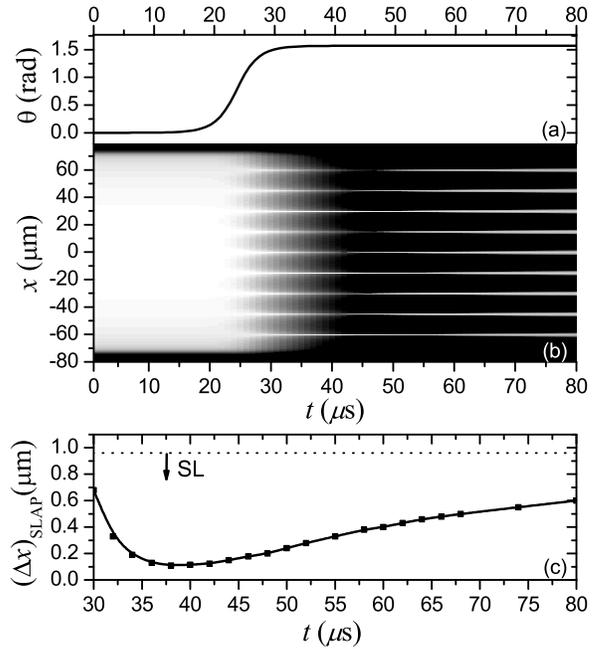}
\caption{SLAP technique for a $^{87}$Rb BEC: (a) Time evolution of the mixing angle $\theta$ at one of the SW anti-nodes. (b) Contour plot of the density distribution of atoms in state $\ket{a}$ as a function of time during the SLAP technique for a SW with 15 $\mu$m period, $\mathcal{R}=100$, $\Omega_{TW0}=2\pi\times 10\cdot 10^6{\rm s}^{-1}$, $\sigma_{TW}=\sigma_{SW}=8\mu$s, $t_{TW}=22\,\mu$s, $t_{SW}=36\,\mu$s, $\Delta_{TW}=\Delta_{SW}=0$,
$\omega_x= 2 \pi \times 14 \,{\rm s}^{-1}$, and $\omega_t= 2 \pi \times 715 \,{\rm s}^{-1}$. (c) Time evolution of the FWHM of the central localized structure.}\label{evolucio}
\end{figure}

To show the time evolution of the system during the SLAP process, we have numerically solved Eqs.~(4)-(6) for a BEC of $5\times 10^4$ atoms. Figs.~5(a) and (b) show the mixing angle $\theta $ at one of the SW anti-nodes and the contour plot of the density distribution of atoms in state $\ket{a}$, respectively, as a function of time and for the parameters given in the figure caption. As expected, component $\ket{a}$ develops extremely narrow structures at the nodes of the SW whose width is much smaller than its spatial period. For demonstration, we have chosen a large spatial period of 15 $\mu$m but arbitrary periods down to $\lambda_{SW} /2$ with a corresponding localization down to the nm scale could be achieved by changing the wave number $k$ of the standing wave. Fig.~5(c) shows the time evolution of the FWHM around one node. The minimal width of the localized structures coincides approximately with the time at which the TW field is switched off, i.e., at $\theta = \pi /2$. We have calculated the transverse momentum spread $\Delta p_x$ for this time, obtaining a beam quality factor \cite{ALas} $M^2 = (2 / \hbar) \Delta x \Delta p_x \approx 0.6$ which is below the Heisenberg limit due to the non-linearities of the two-component trapped BEC partially compensating for diffraction.  

\section{Conclusions and perspectives}

In conclusion, we have introduced the SLAP technique for state-selective localization and patterning of atomic matter waves. We have shown that a 'super-localization' regime beating the previously introduced CPT localization technique \cite{Aga06} can be reached and analytic expressions for the necessary conditions have been derived. We have discussed the use of the SLAP technique for nano-lithograhy with a Ne* matter wave showing the possibility to imprint high contrast patterns with narrow structures whose FWHM approaches the single nanometer scale. This lithographic technique is applicable to all atomic systems with a high-energy dark state formed by the combination of a state fast decaying to the ground and a metastable one. Coherent patterning of a two-component $^{87}$Rb BEC in the super-localization regime has been studied in detail as a second example showing that it is possible to overpass the Heisenberg limit.  

Since localization occurs at the nodes of one of the involved laser fields, more evolved patterning schemes can be realized by extending the present 1D configuration to higher dimensions by applying 2D and 3D SW configurations. Even more complex structures, such as the intensity nodes of higher-order Laguerre-Gauss modes or the light fields of custom-made micro-optical elements \cite{Birkl01}, could be considered for this technique. 
Following the presented SLAP technique for coherent patterning of a BEC, one could consider its application to produce a collection of parallel, coherent, and extremely collimated (pulsed) atom lasers \cite{atomlaser} or, by observing the corresponding near field diffraction pattern, to investigate the matter wave analogue of the optical Talbot effect \cite{Talboteffect}. Finally, the SLAP technique could be also applied to address and detect individual sites in optical lattices by an appropriate choice between the spatial period of the optical lattice and the wavelength of the SW field used in this technique. 

We acknowledge support by the Spanish Ministry of Education and Science under contracts  
FIS2005-01497, FIS2005-01369, FIS2008-02425, HA2005-0002, HD2008-0078, CSD2006-0001,
by the Catalan Government under contract SGR2005-00358, 
by ESF and DFG under the project CIGMA, 
by the European Commission within the RTN Atom Chips and the IP SCALA, 
by the DAAD under contract 0804149,
and by NIST under award 60NANB5D120.

\end{document}